\documentclass{PoS}
\usepackage{amsmath}
\usepackage{slashed}
\usepackage{amssymb}
\usepackage{epsfig}
\usepackage{graphicx}
\usepackage{multirow}
\usepackage{enumitem}
\usepackage[utf8x]{inputenc}
\usepackage{booktabs}
\usepackage{siunitx}
\newcommand{\GeV}{\si{\GeV}}
\newcommand{\TeV}{\si{\TeV}}
\newcommand{\invab}{\si{\per \atto\barn}}
\newcommand{\femtobarn}{\si{\femto\barn}}

\title{Probing light Yukawa couplings in Higgs pair production}
\ShortTitle{Higgs couplings to light quarks via $HH$}
\author{\speaker{Lina Alasfar}\\
{\normalsize Humboldt-Universit\"at zu Berlin, Institut f\"ur Physik,
Newtonstr.~15, 12489 Berlin,  Germany.}\\
        E-mail: \email{lina.alasfar@physik.hu-berlin.de}}
\author{Ramona Gr\"{o}ber\\
{\normalsize Humboldt-Universit\"at zu Berlin, Institut f\"ur Physik,
Newtonstr.~15, 12489 Berlin,  Germany.}\\
{\normalsize Dipartimento di Fisica e Astronomia ``G. Galilei'', Universit\`a di Padova, Italy.}\\
{\normalsize Istituto Nazionale Fisica Nucleare, Sezione di Padova, I-35131 Padova, Italy.}\\
       E-mail: \email{ramona.groeber@pd.infn.it}}
\abstract{One of the puzzles of the SM is the large hierarchy between the Yukawa couplings of different flavours. Yukawa couplings of the first and the second generation are constrained only very weakly so far. However, one can obtain large deviations in the Yukawa couplings in several New Physics (NP) models, such as e.g. models with new vector-like quarks, or new Higgs bosons that couple naturally to individual fermion families. \\ In this work, we investigate the potential bounds on the NP Higgs Yukawa couplings modification $ \kappa_f$ for light quarks from double-Higgs at the LHC, starting from a model independent formalism. We have looked at the two Higgs boson final state $ b \bar b \gamma \gamma $, and the relevant experimental cuts to reduce backgrounds and estimated the potential exclusion bounds for $ \kappa_f$. We have considered both linear and non-linear effective field theory for the Higgs light quark coupling modifications.}
\FullConference{
European Physical Society Conference on High Energy Physics - EPS-HEP2019 -\\
			10-17 July, 2019\\
			Ghent, Belgium \\
		Report Number : HU-EP-19/29.}
\begin{document}
%%%%%%%%%%%%%%%%%%%%%%%%%%%%%%%%%%%%%%%%%%%%
% \section{Introduction}
%%%%%%%%%%%%%%%%%%%%%%%%%%%%%%%%%%%%%%%%%%%%
%
%
%

%%%%%%%%%%%%%%%%%%%%%%%%%%%%%%%%%%%%%%%%%%%%
\section{Effective field theory of light Yukawa couplings}
%%%%%%%%%%%%%%%%%%%%%%%%%%%%%%%%%%%%%%%%%%%%
%
%
%
\par \noindent In the Standard Model~(SM), the Higgs doublet~$\phi$ couples to the quarks via a Yukawa interaction term
\begin{equation}
\mathcal{L}_{y}=-y^u_{ij} \bar{Q}_L^i \tilde{\phi} u_R^j - y^d_{ij} \bar{Q}_L^i \phi d_R^j +h.c.\,,
\end{equation}
with $\tilde{\phi}=i \sigma_2 \phi^*$, $\sigma_2$ is the second Pauli matrix, $Q_L^i$ the left-handed $SU(2)$ quark doublet of the $i$-th generation and $u_R^j$ and $d_R^j$ the right-handed up- and down-type fields of the $j$-th generation, respectively.

\par \noindent It is possible to modify  this coupling via a dimension 6 operator in the SM effective field theory formalism~(SMEFT) with a UV cutoff scale~$\Lambda$
\begin{equation}
\Delta \mathcal{L}_{y}=\frac{\phi^{\dagger}\phi}{\Lambda^2}\left( c^u_{ij} \bar{Q}_L^i \tilde{\phi} u_R^j + c^d_{ij} \bar{Q}_L^i \phi d_R^j +h.c.\right)\,.
\label{eq:EFTop}
\end{equation}
% %
% Since the masses of the quarks are known from measurement, we could write the mass matrices with the new coupling as
% %
% \begin{align}
% M^u_{ij} =& \frac{v}{\sqrt{2}} \left( y^u_{ij}-\frac{1}{2} c^u_{ij}\frac{v^2}{\Lambda^2}\right)\,,\\
% M^d_{ij} =& \frac{v}{\sqrt{2}} \left( y^d_{ij}-\frac{1}{2} c^d_{ij}\frac{v^2}{\Lambda^2}\right)\,.
% \end{align}
% %
The resulting mass matrices can be diagonalised by means of a bi-unitary transformation~$V_{L/R}^{q}$ in order to recover the measured quark masses.  The Yukawa couplings~($h \bar q q$) will be modified by this new operator. Moreover, a new coupling between quark anti-quark pair and two Higgs bosons $hh \bar q q$ appears  with coupling constant
\begin{equation}
g_{h\bar{q}_i q_j} : \quad \frac{m_{q_i}}{v}\delta_{ij}-\frac{v^2}{\Lambda^2} \frac{\tilde{c}^q_{ij}}{\sqrt{2}}\,, \quad \quad \quad \quad \quad g_{h h\bar{q}_i q_j} : \quad -\frac{3}{2\sqrt{2}}\frac{v}{\Lambda^2}\tilde{c}^q_{ij}\,, \label{eq:couplingsEFT}
\end{equation}
where
\begin{equation}
\tilde{c}^{q}_{ij}= \left(V_{L}^{q}\right)^*_{ni}c_{nm}^{q}\left(V_R^{q}\right)_{mj}\, , \; \; \; \;  \text{with } \;\;\;\;\; q = u,d\,,
\end{equation}
and $V_{L/R}^{q}$ denote the transformation matrices from current to mass eigenstates.
\par \noindent It has been shown in~\cite{Bar-Shalom:2018rjs,Egana-Ugrinovic:2018znw,Alasfar:2019pmn} that it is possible to introduce large modifications to the Yukawa couplings without flavour-changing neutral currents (FCNCs) at tree-level, which strongly constrain off-diagonal couplings~\cite{Blankenburg:2012ex}, and without assuming minimal flavour violation (MFV)~\cite{DAmbrosio:2002vsn}.
Very little is known though about the flavour-diagonal Higgs coupling to light quarks, as seen from the weak bounds on these coupling modifications, \textit{cf.}~ref.~\cite{deBlas:2019rxi}.
Therefore, we only restrict ourselves to varying the diagonal light quark Higgs couplings, writing
\begin{equation}
g_{h\bar{q}_i q_i} =\kappa_q g_{h\bar{q}_i q_i}^{\text{SM}} \,, \quad \quad \quad \quad \quad g_{h h\bar{q}_i q_i}= - \frac{3}{2}\frac{1-\kappa_q}{v}g_{h\bar{q}_i q_i}^{\text{SM}} \,.
\label{eq:def_kappa}
\end{equation}
Here, we have abused the~$\kappa$-formalism notation that is used heavily in experimental searches.
\par \noindent Another way to modify the Yukawa couplings  in a model-independent way is via the electroweak chiral Lagrangian~\cite{Coleman:1969sm}.
Using the chiral Lagrangian by taking the 0th chiral mode and restricting ourselves to the flavour diagonal  couplings we get
\begin{equation}
    -\mathcal L = \bar q_{L} \frac{m_q}{v} \left( v + c_{q}h + \frac{c_{qq}}{v} h^2 + \dots \right) q_R + h.c.
		\label{chirallag}
\end{equation}

Thus, in this formalism, the Yukawa and $hh \bar q q$ couplings are uncorrelated, making the Higgs pair production the most accessible process to test the potential correlation between the Higgs-light quark couplings.
% removed due to lack of allowed space
% \par \noindent Figure~\ref{fig_uv_qqhh}illustrates examples of concrete models that could contribute to the dim 6 operator in eq~\eqref{eq:EFTop}.
% %%%
% \begin{figure}[!t]
% \centering
%   \includegraphics[width = 0.25\textwidth]{qqh_2hdm}
%    \hspace{0.5 cm}
%   \includegraphics[width = 0.2\textwidth]{VLQ}
%   \caption{Examples of concrete models leading to a  $hh q \bar{q} $ coupling. The left Feynman diagram shows a heavy Higgs $H$, the right diagram a vector-like quark $\mathcal Q $.}
%   \label{fig_uv_qqhh}
% \end{figure}
% %%%
% For a heavy scalar~$H$~(e.g. 2 Higgs doublet Model) we cold match the  $hh q \bar{q} $  with
% \begin{equation}
% g_{hh\bar{q}q} \to -i \frac{g_{H\bar{q}q} g_{Hhh}}{m_H^2}\,.
% \end{equation}
% Alternatively, if the $hh q \bar{q} $ coupling emerges from a concrete model with vector like quarks~${\cal Q}$ we could do the matching
% \begin{equation}
% g_{hh\bar{q}q} \to -i \frac{g_{h\bar{q}\mathcal Q}g_{h\bar{\mathcal Q}q}}{m_{ \cal Q}}\,.
% \end{equation}

%%%%%%%%%%%%%%%%%%%%%%%%%%%%%%%%%%%%%%%%%%%%
\section{Higgs pair production with modified Yukawa couplings}
%%%%%%%%%%%%%%%%%%%%%%%%%%%%%%%%%%%%%%%%%%%%
%
%
%
\par \noindent In the SM, the production of Higgs pairs in hadron colliders~(such as the LHC) is dominated by gluon-gluon fusion~(ggF) which at leading order~(LO) is $ \sim 21 \femtobarn$.
% , as illustrated by the Feynman diagrams in fig.~\ref{fig_ggf_sm}.
% %
% \begin{figure}[!t]
% \centering
%   \includegraphics[width = 0.35\textwidth]{ggfbox}
%  \hspace{0.3 cm}
%   \includegraphics[width = 0.4\textwidth]{ggftri}
%   \caption{Feynman diagrams for the ggF process of Higgs pair production in the SM.}
%   \label{fig_ggf_sm}
% \end{figure} to this process have been computed up to order alphas^4
% %

\begin{figure}[!tb]
\centering
\begin{picture}(180,200)
\put(-120,120){\includegraphics[scale =0.25]{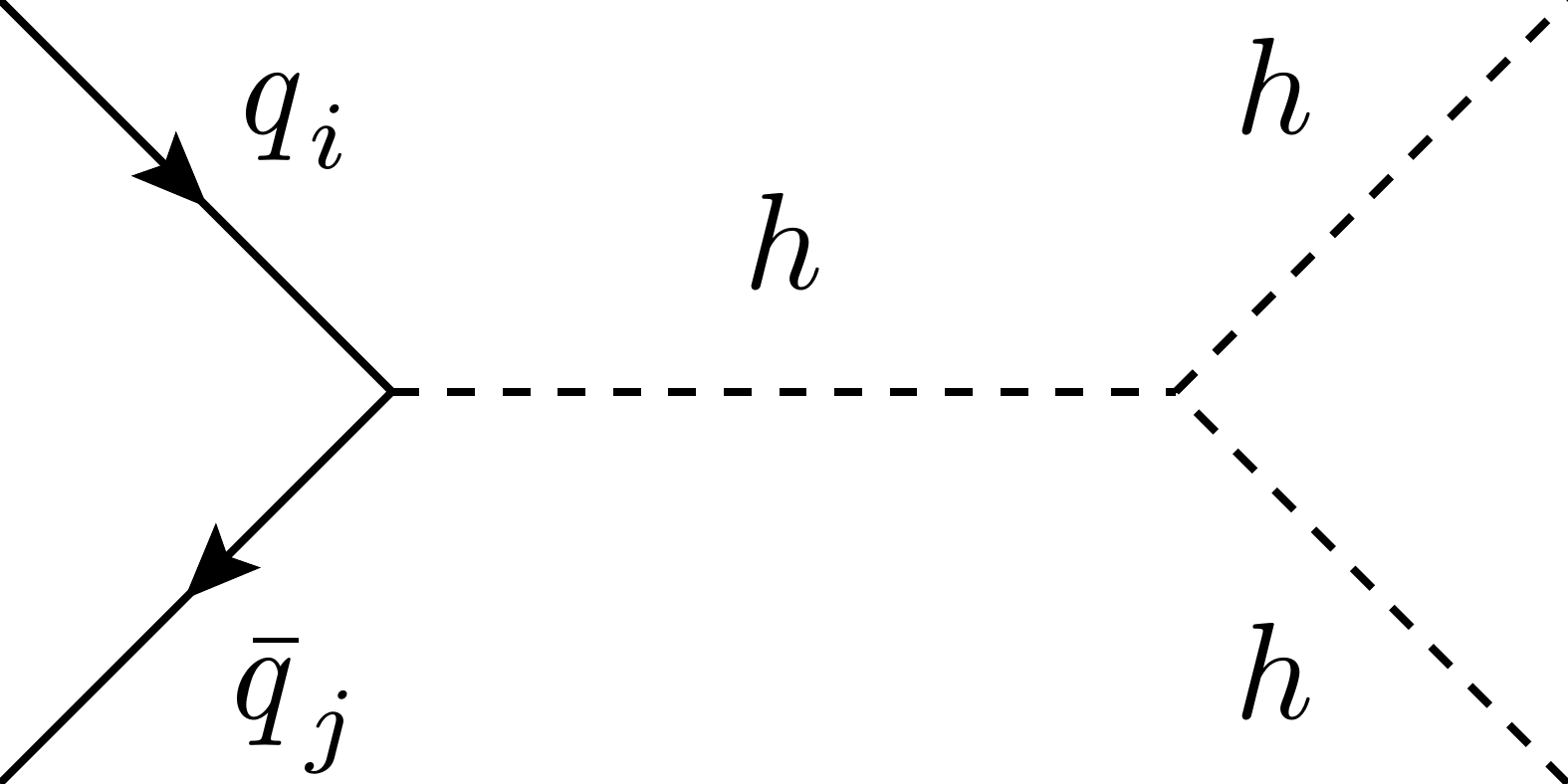}}
\put(20,120){\includegraphics[scale = 0.25]{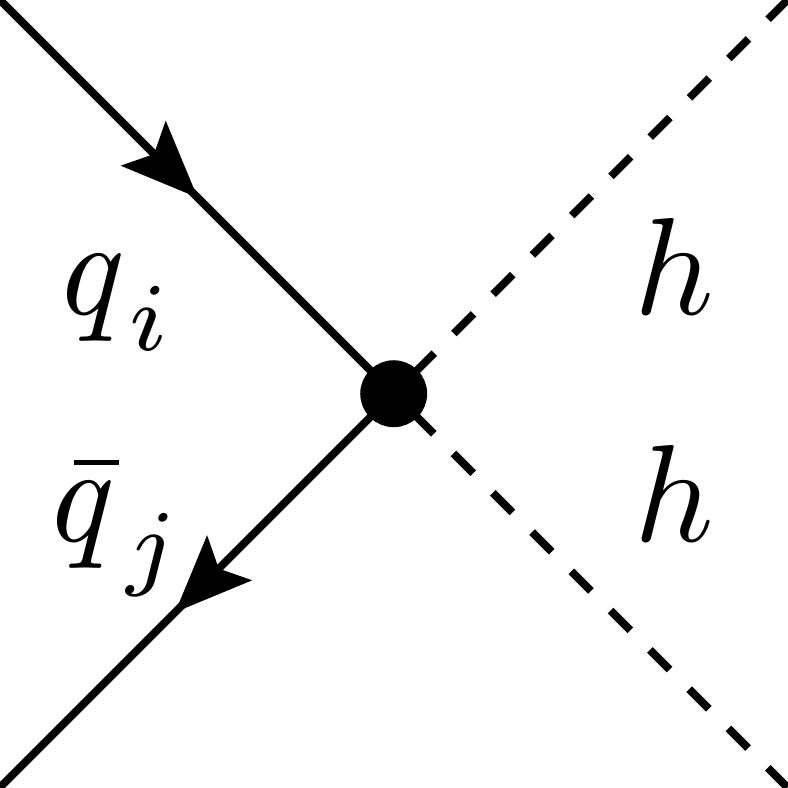}}
\put(110,110){\includegraphics[scale =0.21]{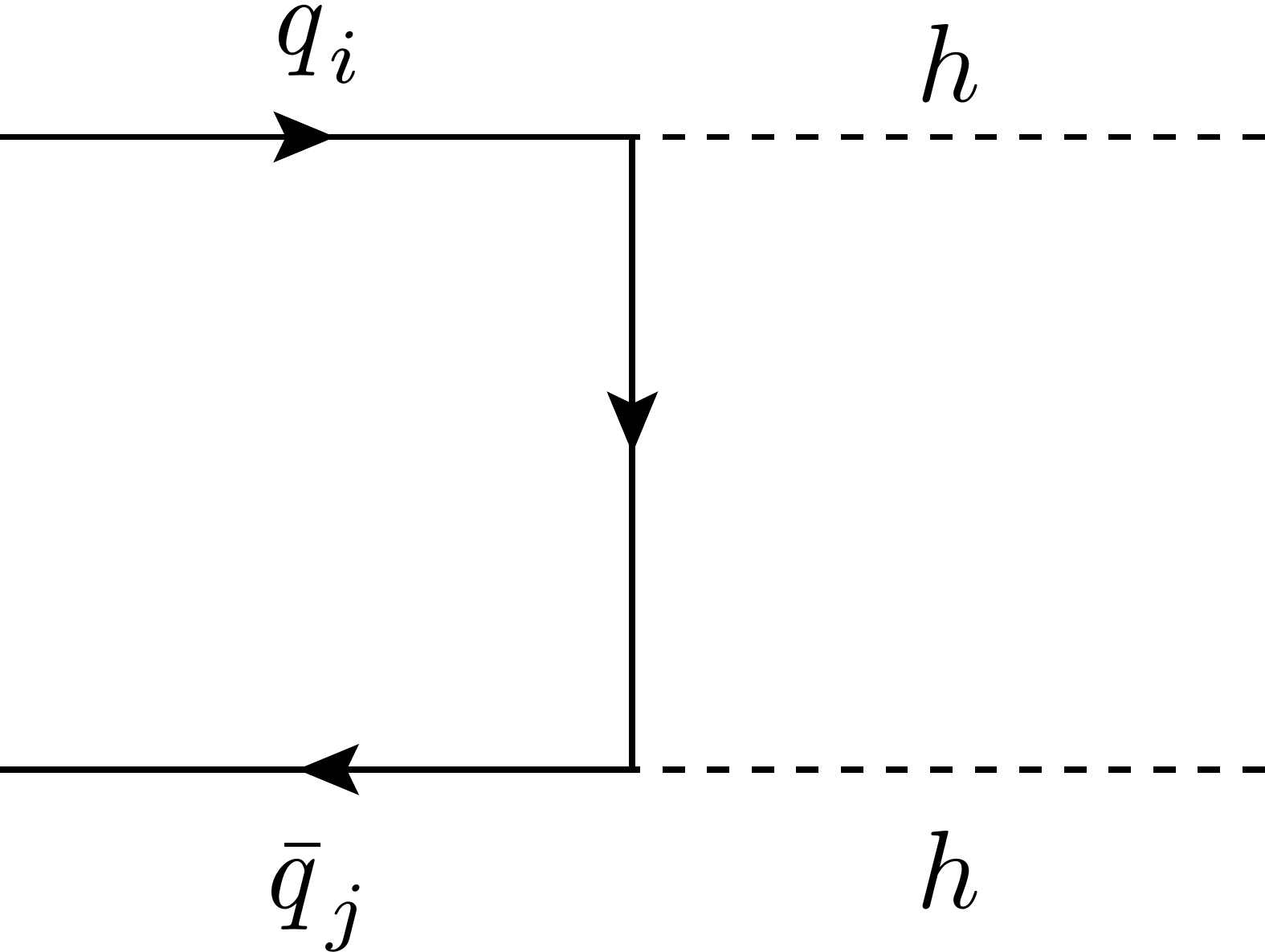}}
\put(230,145 ){{\large+ crossed} }
\end{picture}
\vspace*{-4cm}
  \caption{ Feynman diagrams for the qqA Higgs pair production.}
  \label{qqA_fd}
\end{figure}

Higher order QCD corrections to this process have been computed up to order 
$\alpha_s^4$~\cite{EBOLI1987269}.  For the inclusive cross section the LO calculation can be corrected by a K-factor,
\begin{equation}
  K = \frac{\sigma_{NNLO}}{\sigma_{LO}}, \;\;\;\;\; K_{14\, \mathrm{ TeV}} = 1.72.
\end{equation}
In our analysis, we use the state-of-the-art calculation of the ggF inclusive cross section at 14 \TeV~\cite{Grazzini:2018bsd} $\sigma^{\text{SM}}_{NNLO}=36.69^{+1.99}_{-2.57} \text{ fb}\,.
$,
%
% \begin{equation}
% \end{equation}
%

%
However, if the light quark Yukawa couplings are significantly increased, the ggF process will become subdominant compared to the quark anti-quark annihilation~(qqA), due to the abundance of light quarks in the proton. This happens if the coupling modifiers are of order   $\kappa_c^{qqA = ggF} \sim 5$, $\kappa_s^{qqA = ggF} \sim 10$ and $\kappa_u^{qqA = ggF} \sim \kappa_d^{qqA = ggF} \sim 10^3$. Those values are not yet excluded by current measurements. 
The qqA Higgs pair production is calculated from the diagrams shown in fig.~\ref{qqA_fd}. up to NLO QCD, see~\cite{Alasfar:2019pmn}.
% \Lina{This part could be removed}\\
% \par \noindent Choosing a benchmark point for the light Yukawa scalings where all the light couplings are scaled to the $b$-quark SM Yukawa  \footnote{This corresponds to the the scalings of values: \[\kappa_u=  1879\,, \hspace*{0.5cm} \kappa_d= 889 \,, \hspace*{0.5cm}\kappa_s= 44\,, \hspace*{0.5cm} \kappa_c =3.3\,, \label{eq:fitbounds}\]} one finds the NLO qqA cross section to be
% %
% \begin{equation}
% \sigma^{qqA}_{NLO}= 284 \pm 25 \text{ fb}\,,
% \end{equation}
% could  be included or not
% \par \noindent When comparing the `normalised' differential cross section distributions of both ggF from SM and qqA at our benchmark point in figure~\ref{qqA_dsigdmhh}, a significant shape difference is seen. The qqA has a shifted to the left and less defined peak.
% \begin{figure}[!b]
% \centering
%   \includegraphics[width = 0.7\textwidth]{uh_shnnlo_shape.pdf}
%   \caption{The qqA normalised NLO invariant mass differential cross section distribution for the benchmark point ($g_{hq \bar q} = g_{h b \bar b}^{\SM}$) (solid line) and the NNLO SM ggF cross section obtained from~\cite{Grazzini:2018bsd} (dashed line).
%   }
%   \label{qqA_dsigdmhh}
% \end{figure}
The shape difference between the normalized ggF and qqA dominant distributions is used in the analysis to extract sensitivity limits on modifications of light quark Yukawa couplings.
%%%%%%%%%%%%%%%%%%%%%%%%%%%%%%%%%%%%%%%%%%%%
\section{Phenomenological analysis}
%%%%%%%%%%%%%%%%%%%%%%%%%%%%%%%%%%%%%%%%%%%%
%
%
%
\par \noindent In order to estimate the sensitivity of the  high-luminosity LHC~(HL-LHC) for the light quark scalings, we have used the profile likelihood method to estimate the signal strength~$\mu$, defined as~(for general resonance $R$ production)
\begin{equation}
    \mu = \frac{ N_{expec}}{ N^{SM}_{expec}}, \;\;\;\;\; \text{with} \;\;\;\;  N_{expec} = \sigma(pp\to R) \, \mathcal B(R \to X)\, L \, \epsilon_{\mathrm{SEL}},
\end{equation}
% with
% \begin{equation}
%
% \end{equation}
where $ \sigma(pp\to R)$  $ \, \mathcal B(R \to X)$ is the production cross section of $R$ times the branching fraction for the final state $X$, $L = 3\invab$ the integrated luminosity of the HL-LHC, and $ \epsilon_{\mathrm{SEL}}$ is the experimental selection.
\par \noindent In our analysis we have chosen the two Higgs final state~$hh\to b \bar b \gamma \gamma$ due to its high potential of reconstruction at the HL-LHC~\cite{  Baur:2003gp, Baglio:2012np}. We have modified the FORTRAN programme \texttt{HDECAY}~\cite{Djouadi:1997yw} to obtain the Higgs boson branching ratios, including state-of-the art QCD corrections and the light fermion loops and decay channels.
\par \noindent The Higgs pair production and decays were implemented into \texttt{Pythia} 6.4 for parton showering. Then the events were analysed according
to~\cite{Azatov:2015oxa} in order to estimate the selection efficiency.
In addition to $b$-tagging, we have used charm mistagging probability of $b$-jets combined with $c$-tagging working points for CMS~\cite{Chatrchyan:2013zna} and ATLAS~\cite{Aad:2014xzb}, as developed in~\cite{Perez:2015aoa,Kim:2015oua} in order to constrain $\kappa_c$ and $\kappa_s$.
This will allow us to put bounds on the second generation quark Yukawa couplings, while otherwise no sensitivity can be obtained, due to the fact that the lowering of the branching fraction of the final state~$hh\to b \bar b \gamma \gamma$ is larger than the cross-section enhancement. This method will allow us to probe the final state ~$hh\to c \bar c \gamma \gamma $ using the current $c$-tagging techniques \cite{Aad:2015gna,ATLAS-CONF-2013-063} or the improved ones after the ATLAS HL-LHC upgrade~\cite{Capeans:1291633,ATL-PHYS-PUB-2015-018}, thus in total improving the expected sensitivity for the second generation couplings.
%%%%

%%%%%%%%%%%%%%%%%%%%%%%%%%%%%%%%%%%%%%%%%%%%
\section{Results}
%%%%%%%%%%%%%%%%%%%%%%%%%%%%%%%%%%%%%%%%%%%%
%
\par
\begin{figure}[!htbp]
\centering
  \includegraphics[width = 0.65\textwidth]{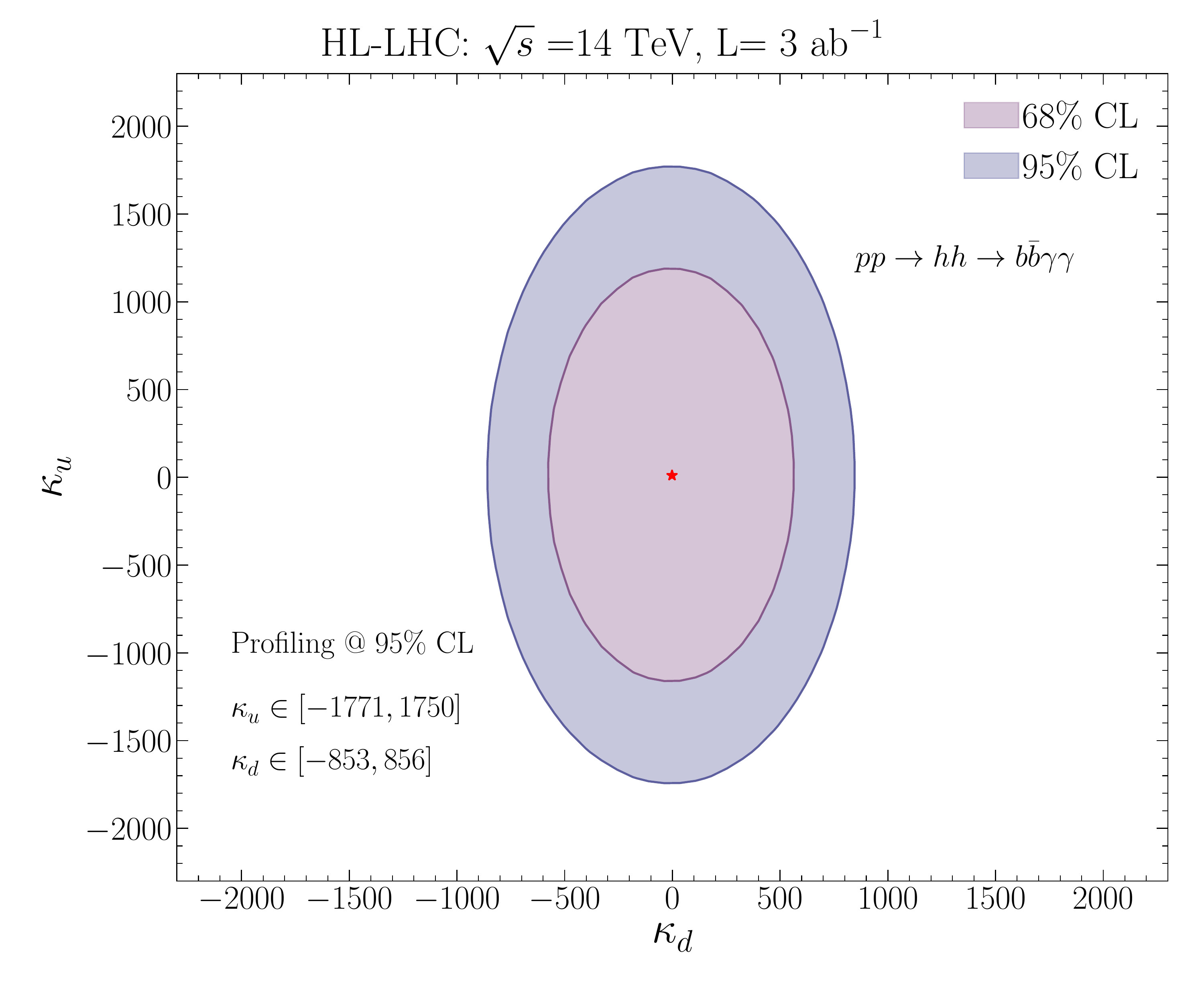}
  \caption{The expected sensitivity likelihood contours  at  68\% and 95\% CL of the HL-LHC for the first generation Yukawa coupling scalings. }
  \label{bounds_1stgen}
\end{figure}

 \noindent Using a likelihood fit for estimating $\mu$, assuming that SM Higgs pair production is the null hypothesis, then preforming a scan over $ \kappa_u$ and $ \kappa_d$, we obtain the 68\% and 95\% CL sensitivity likelihood contours for the HL-LHC shown in fig.~\ref{bounds_1stgen}.
%
% Profiling the 2D likelihood allows us to get the HL-LHC sensitivity upper bound for the individual first generation scalings :
% \begin{equation}
%    -571  < \kappa_d <  575, \;(\text{68\% CL}), \, \;\;\;\; \,
%      -853 < \kappa_d <  856,\;(\text{95\% CL}),
% \end{equation}
%  and %
% \begin{equation}
%    -1192  < \kappa_u < 1170, \;(\text{68\% CL}), \, \;\;\;\; \,
%     -1771 < \kappa_u <  1750,\;(\text{95\% CL}).
% \end{equation}
% %

\par
\begin{figure}[!htpb]
\centering
 \includegraphics[width = 0.45\textwidth]{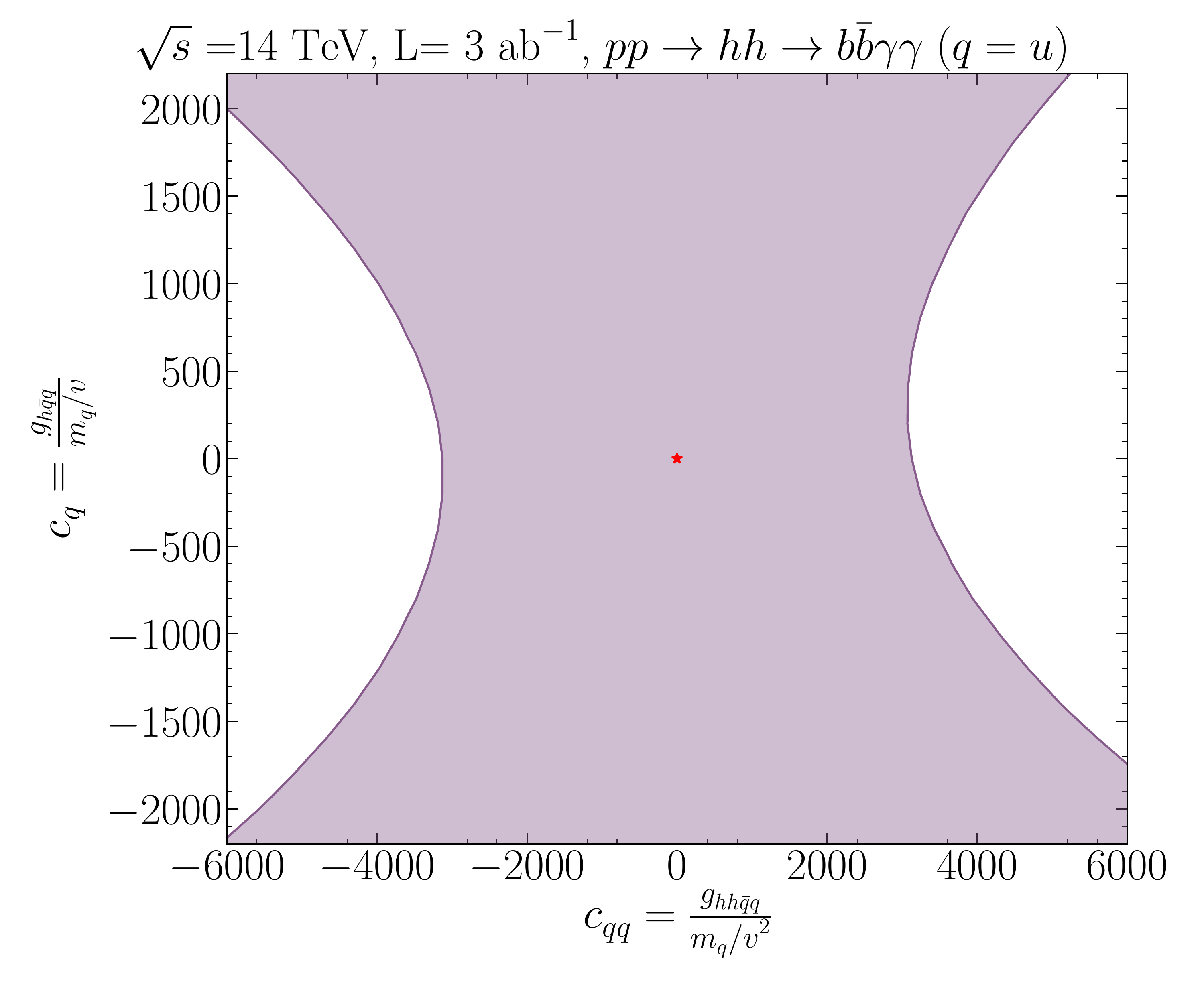}
 \includegraphics[width = 0.45\textwidth]{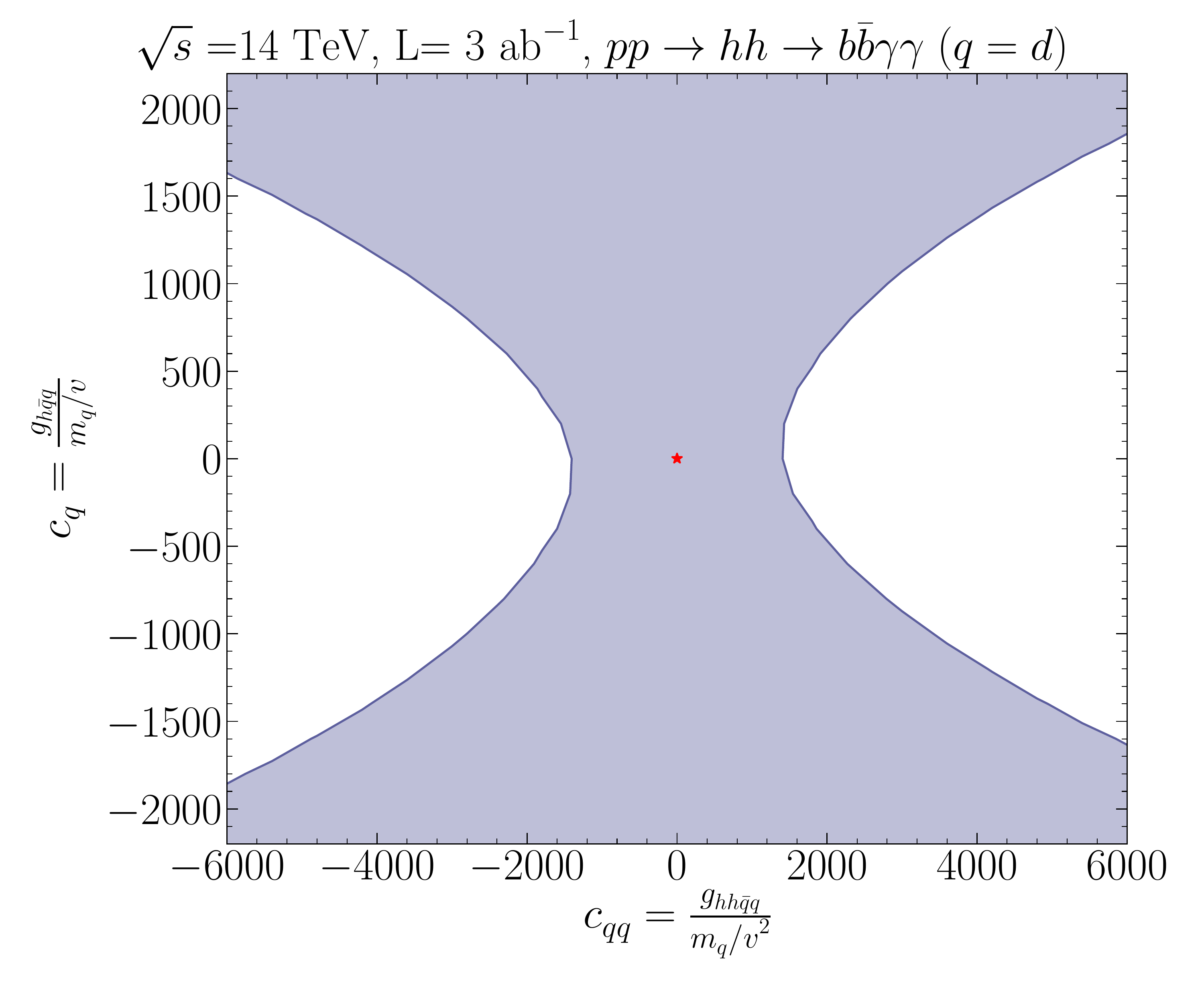}
 \\
 \includegraphics[width = 0.45\textwidth]{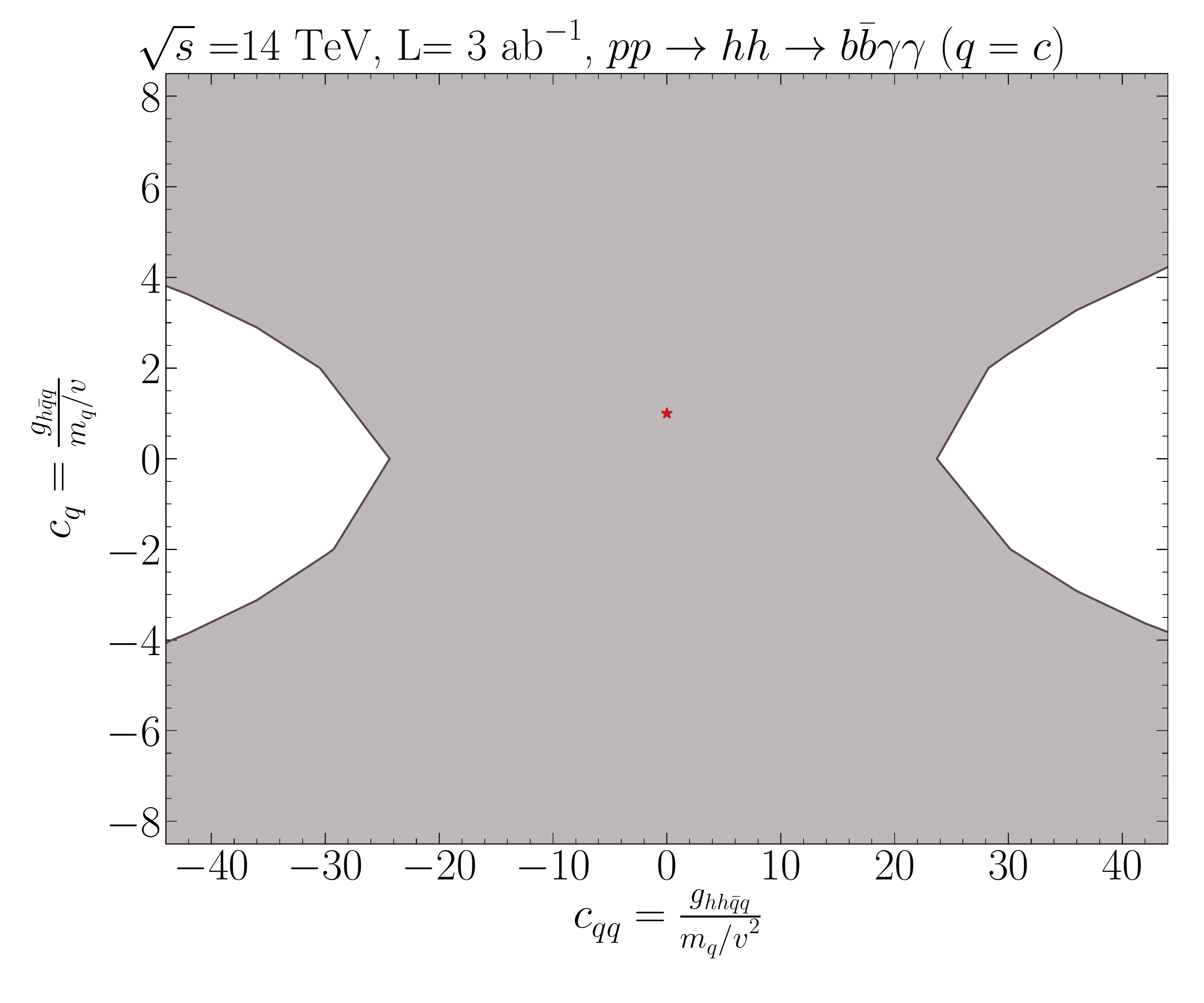}
 \includegraphics[width = 0.45\textwidth]{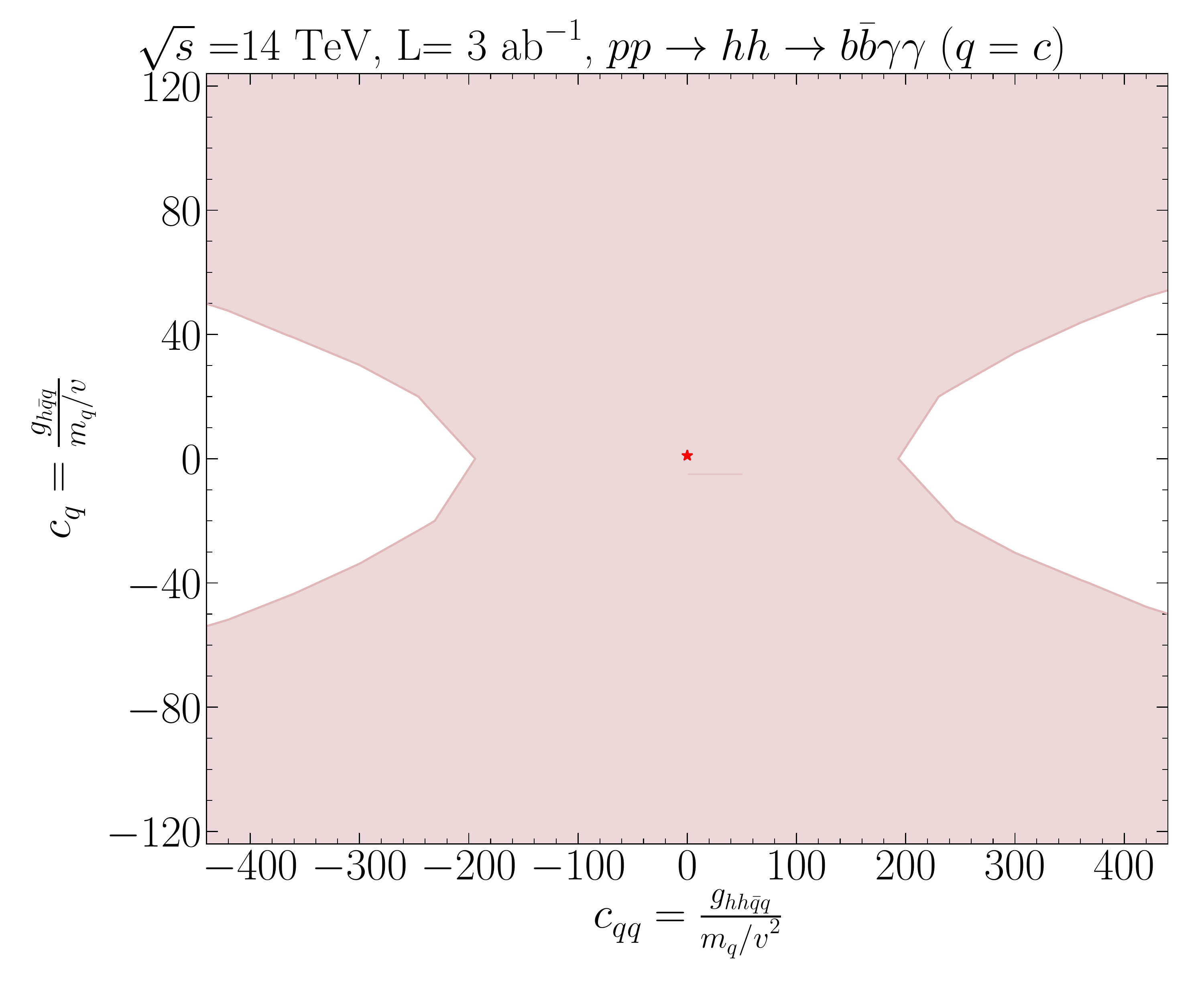}
 \caption{95\% CL likelihood contours for the non-linear EFT  coefficients $c_{qq}$ and $ c_{q}$ for up  (\textit{upper left}), down  (\textit{upper right}), charm (\textit{lower left}) and strange quarks (\textit{lower right}).}
 \label{nleft}
\end{figure}

 \noindent We show now results for non-linear EFT Wilson coefficients $c_{q}$ and $c_{qq}$ of the chiral Lagrangian of eq.~\eqref{chirallag} by
 scanning over them separately in order to obtain the sensitivity bounds on the non-linear EFT illustrated in fig.~\ref{nleft}.
\par
\begin{figure}[!htpb]
\centering
  \includegraphics[width = 0.45\textwidth]{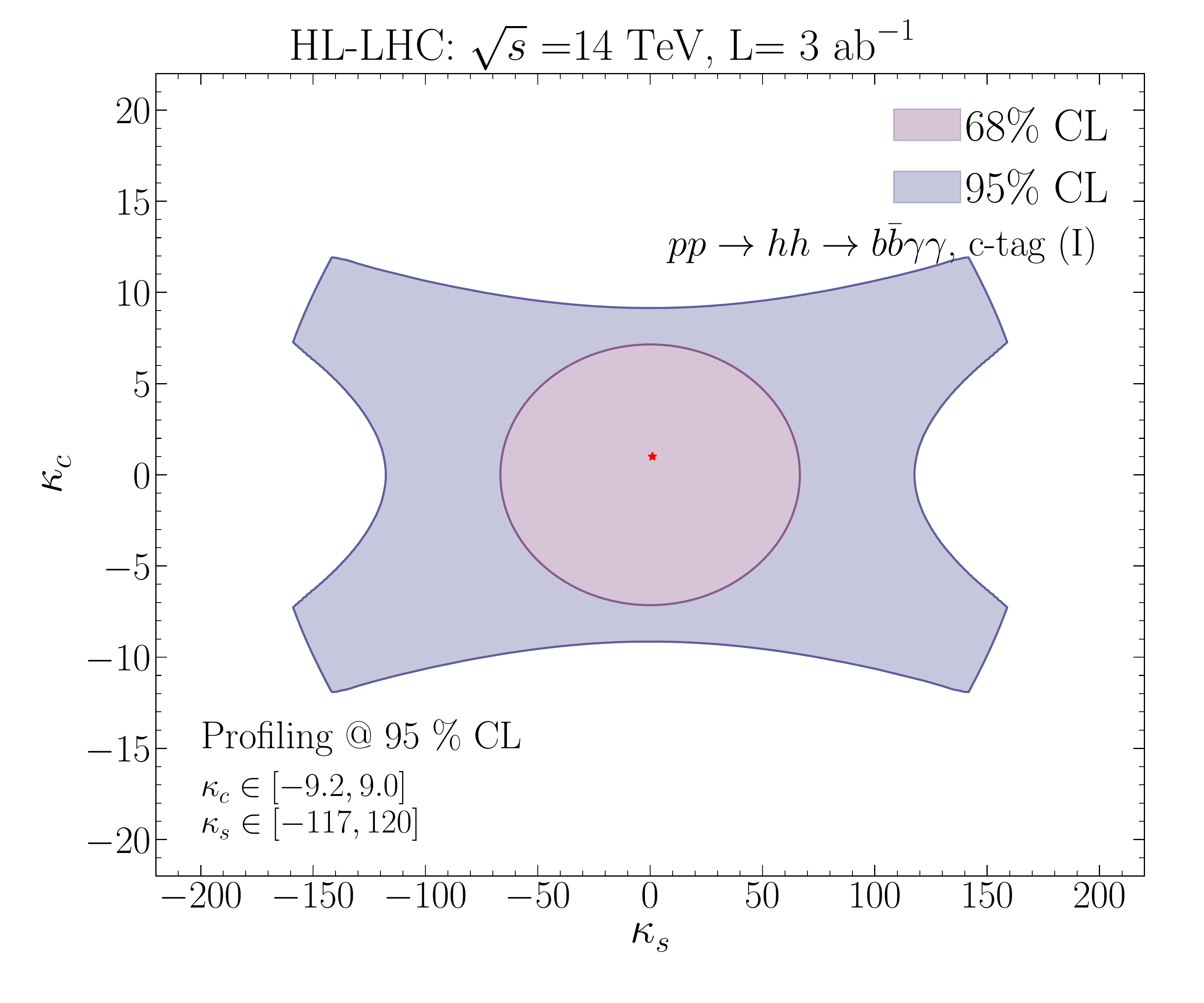}
   \includegraphics[width = 0.45\textwidth]{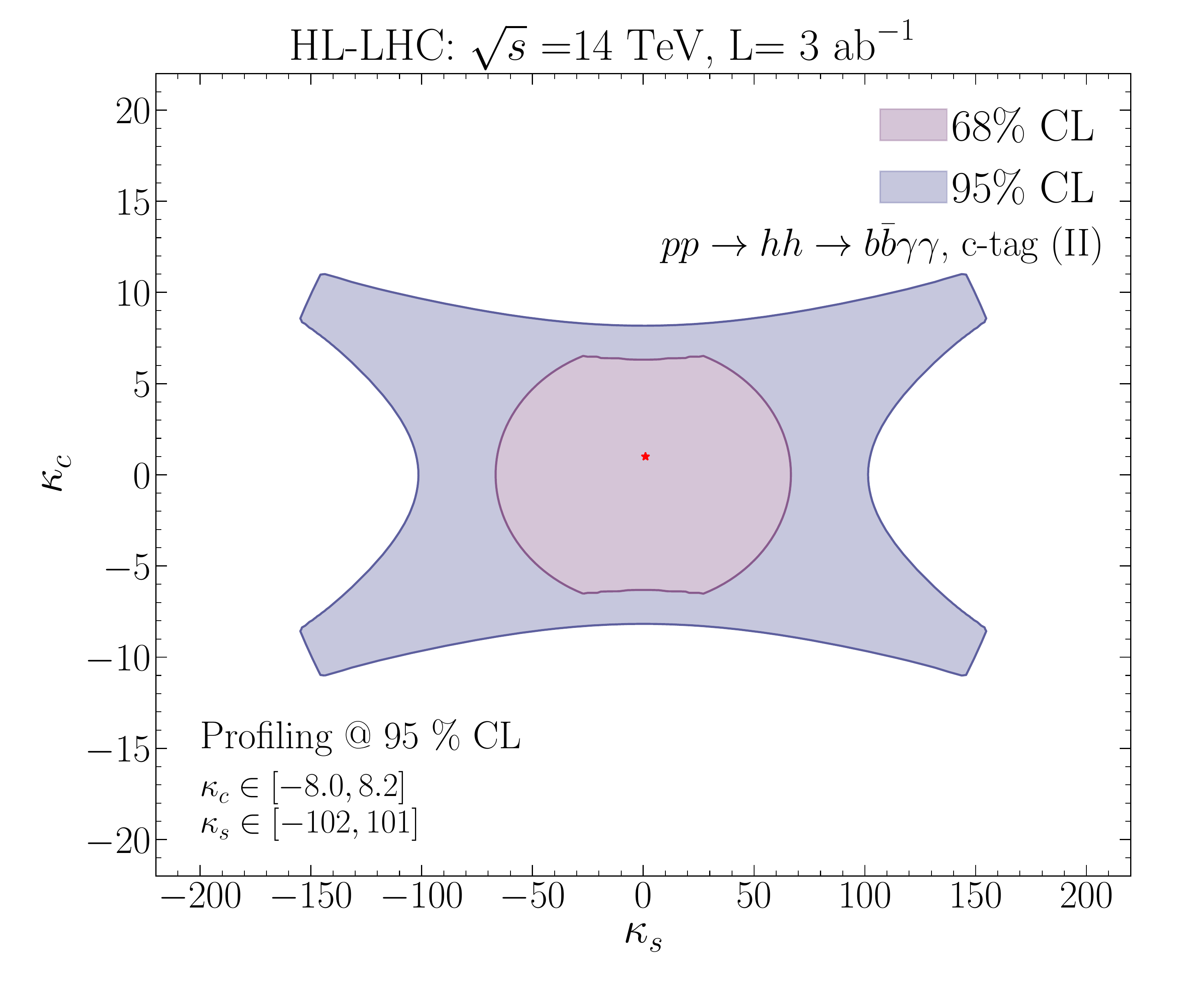}
   \centering
   \includegraphics[width = 0.45\textwidth]{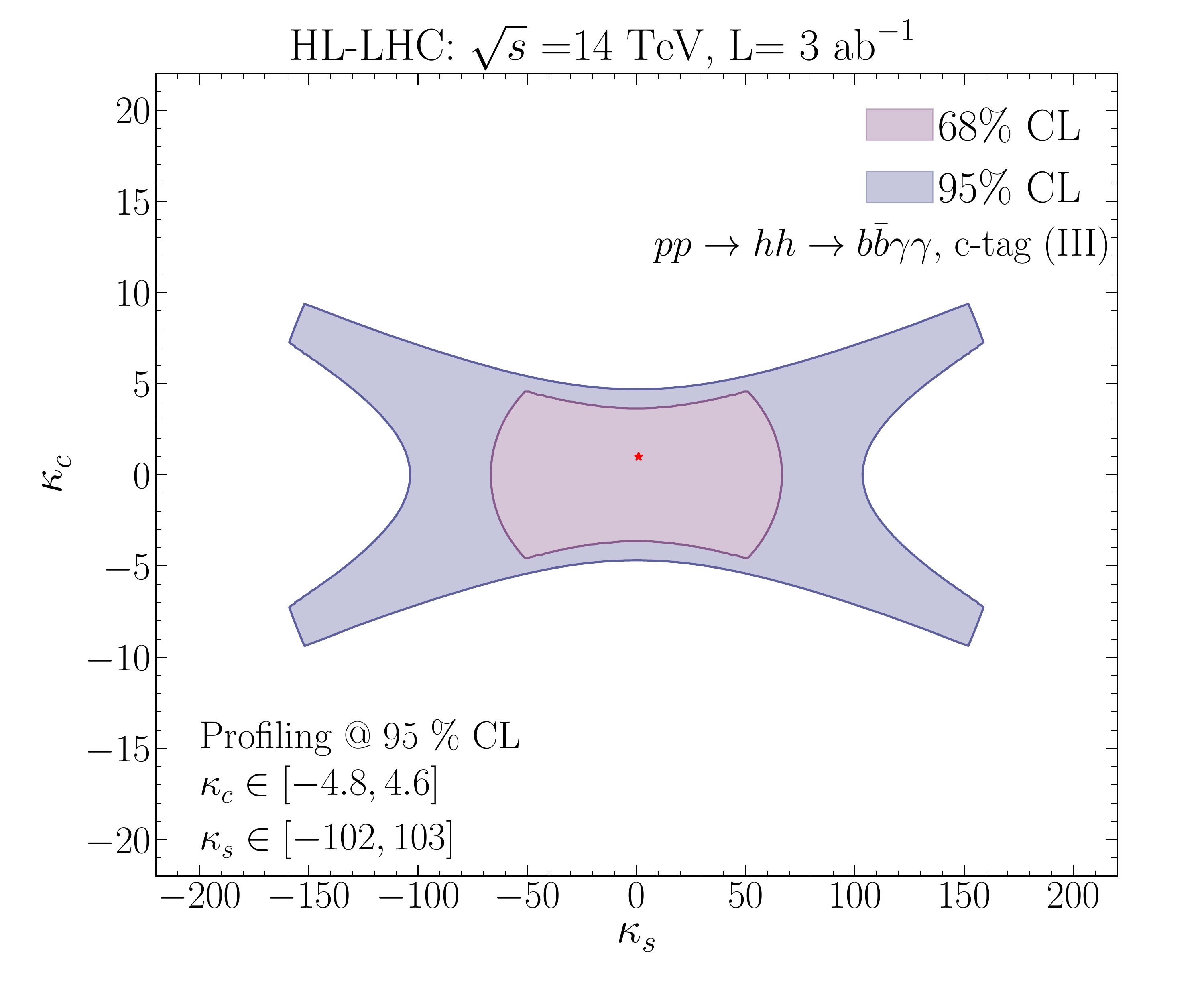}
  \caption{The expected sensitivity likelihood contours at 68\% CL and 95\% CL for an integrated luminosity $L=3 \invab$ for modified second generation quark Yukawa couplings, using the c-tagging I (\textit{upper pannel, left}), II (\textit{upper pannel, right}) and III (\textit{lower pannel}) working points, see~\cite{Alasfar:2019pmn}. }
  \label{bounds_2ndgen1}
\end{figure}

 \noindent For the linear EFT hypothesis, it was not possible to construct sensitivity bounds using only the final state with $b$-quarks for the second generation quarks as discussed before. We hence directly include into fig.~\ref{bounds_2ndgen1} final states with $c$-quarks, which depends on the $c$-tagging working points as discussed also in~\cite{Alasfar:2019pmn, Perez:2015aoa,Perez:2015lra}.
The expected sensitivity limits for the charm quark Yukawa modifications are an improvement compared to the current direct bounds and prospects for the HL-LHC, see~\cite{Perez:2015aoa,Perez:2015lra}.
% \begin{figure}[!t]
% \centering
%   \includegraphics[width = 0.75\textwidth]{ll_contours_kfklambda}
%   \caption{The expected sensitivity likelihood contours at 95\% CL for an integrated luminosity $L=3000\text{ fb}^{-1}$ for modified Higgs trilinear coupling $\kappa_\lambda$ vs the light quark Yukawa couplings scalings $\kappa_q$.}
%   \label{klkf}
% \end{figure}

%%%%%%%%%%%%%%%%%%%%%%%%%%%%%%%%%%%%%%%%%%%%
\section{Conclusion}
%%%%%%%%%%%%%%%%%%%%%%%%%%%%%%%%%%%%%%%%%%%%
%
%
%
\par \noindent Searching for the production of Higgs pairs ($hh$) provides a great insight into the least understood Higgs couplings, the trilinear Higgs self-coupling and the light quark Yukawa couplings. As we showed, it is possible to set prospective model-independent bounds on the light quark Yukawa couplings which are comparable to the prospects from a global fit making use of the Higgs pair production process.  The expected sensitivity is$|\kappa_u| \lesssim 1170 $ and $|\kappa_d| \lesssim 850 $, \textit{cf}.~fig.~\ref{bounds_1stgen}.
\par \noindent Moreover, with mixed $b$- and $c$-tagging working points, the Higgs pair production provides  competitive sensitivity for constraining modifications of the charm Yukawa coupling i.e. $|\kappa_c| \lesssim 5$ and $|\kappa_s | \lesssim 100 $, \textit{cf.}~fig.~\ref{bounds_2ndgen1}, where the first prospective limit is comparable to the prospects from charm tagging in the $Vh$ channel~\cite{Perez:2015aoa}. Furthermore, it turns out that the process is in particularly sensitive to the non-linearities in the light quark-Higgs couplings, so providing a possible insight into the linear or non-linear nature of the Higgs boson.

%%%%%%%%%%%%%%%%%%%%%%%%%%%%%%%%%%%%%%%%%%%%%%
\section{Acknowledgements}
We thank R.~Corral Lopez for collaboration on this topic and the organisers of the conference for a pleasant and stimulating atmosphere during the conference. RG was in parts supported by the ``Berliner Chancengleichheitsprogramm''.
%
%\bibliographystyle{utphys.bst}
%\bibliography{bibliography}

\end{document}